# Hyperbolic Dispersion and Low-Frequency Plasmons in Electrides

Qi-Dong. Hao[1,2], Hao. Wang[1], Hong-Xing. Song[1], Xiang-Rong. Chen[2]\*, Hua-Yun. Geng[1,3]\*

[1] *National Key Laboratory of Shock Wave and Detonation Physics, Institute of Fluid Physics, China Academy of Engineering Physics, Mianyang, Sichuan 621900, P. R. China;*

[2] *College of Physics, Sichuan University, Chengdu 610065, P. R. China;*

[3] *HEDPS, Center for Applied Physics and Technology, and College of Engineering, Peking University, Beijing 100871, P. R. China.*

**Abstract:** Natural hyperbolic materials have attracted significant interest in the field of photonics due to their unique optical properties. Based on the initial successful explorations on layered crystalline materials, hyperbolic dispersion was associated with extreme structural anisotropy, despite the rarity of natural materials exhibiting this property. Here we show that non-cubic electrides are generally promising natural hyperbolic materials owing to charge localization in interstitial sites. This includes elemental and binary electrides, as well as some two-dimensional materials that show prominent in-plane hyperbolic dispersion. They exhibit low plasma frequencies and a broad hyperbolic window spanning the infrared to the ultraviolet. In semiconductor electrides, anisotropic interband transitions provide an additional mechanism for hyperbolic behaviour. These findings remove the previously held prerequisite of structural anisotropy for natural hyperbolic materials, and open up new opportunities, which might change the current strategy for searching and design photonic materials.

## Introduction

The quest for hyperbolic materials has profoundly impacted the field of photonics, showcasing unique optical properties such as support for ultra-high wave vectors[1,2], subwavelength imaging capabilities[3,4], and full-angle negative refraction[5]. These characteristics stem from the highly anisotropic nature of their dielectric tensor, where the sign of the real part of the dielectric function differs along different axes within the interested frequency ranges. This unique property leads to hyperbolic dispersion, creating hyperboloid-shaped constant frequency surfaces in momentum space, in contrast to the ellipsoidal shapes typical of isotropic media[6]. Hyperbolic dispersion is typically realized in artificial hyperbolic metamaterials[7] or in certain natural hyperbolic materials (NHMs). The latter has become especially attractive in recent years due to their atomic-scale structural periodicity and the of that without the requirement for complex nanostructuring techniques[8,9].

\* *Corresponding authors. E-mail: s102genghy@caep.cn; xrchen@scu.edu.cn*

So far, most reported NHMs are in a layered crystalline structure, where the material consists of well-defined atomic layers separated by van der Waals gaps or coupled by weak interlayer interactions. The intrinsic layering naturally leads to structural anisotropy between the in-plane and out-of-plane directions, thus favors the emergence of hyperbolic behavior in plasmons and phonon-polaritons along these two directions[10-14]. Additionally, some highly anisotropic biaxial crystals can also exhibit hyperbolic polaritons with narrow spectral windows, even without relying on layered structures. These are mainly strongly localized phonon-polaritons, such as those in α-$MoO_3$[15,16], $Al_2O_3$[17], and α-$V_2O_5$[18]. In contrast, in-plane hyperbolic plasmons and hyperbolic exciton-polaritons that do not depend on layered structures are exceedingly rare, and have only been experimentally confirmed in $WTe_2$[19], $CrSBr$[20], and black phosphorus [21]. For naturally occurring crystals with moderate anisotropy and non-layered structure—such as hexagonal and tetragonal systems—no significant hyperbolic dispersion has been reported. This led to the general belief that strong structural anisotropy is a prerequisite for hyperbolic dispersion to occur. Driven by this faith, the exploration of NHMs focused on highly anisotropic structures in recent years, despite the fact that such materials are extremely rare. This has greatly limited the progress in searching for NHMs with improved properties.

On the other hand, recent studies have revealed a large number of extraordinary phases of materials, including exotic electrides where the excessive charge is localized into the interstitial sites (forming interstitial quasi-atom or ISQ)[22,23]. As an emerging class of materials, electrides have been reported in a wide range of elements that span from simple metals to complex compounds [24-27]. They might harbor unique physical properties arising from the nuclei charge polarization[28-30], as well as strong electron-phonon coupling[31] and extraordinary optical properties[32]. The latter is also due to the interaction between ISQs and nuclei in electrides that not only modulating the electronic band structure, but also diverging the optical anisotropy. That is, many non-cubic electrides might be potential candidates for NHMs, besides the known layered structures. Previous studies have demonstrated that layered electrides such as $Ca_2N$ and $Sr_2N$ exhibit significant hyperbolic dispersion[33,34]. However, it is unclear whether the charge localization contributes significantly to the formation of hyperbolic dispersion or not, since layered structure alone also would yield hyperbolic dispersion irrespective of the charge localization.

In this work, we discover that charge localization and the resultant polarization in electrides significantly impact the optical response, lowering the plasma frequency and inducing large dielectric anisotropy. As a result, the vast majority of non-cubic electrides are potentially excellent natural hyperbolic materials. We illustrate this by showing that the natural elemental metallic electride Be under ambient conditions has bulk plasma frequencies of only 1.06 eV (along *a* and *b* directions) and 1.94 eV (along *c* direction), respectively, which leads to prominent hyperbolic plasmons ranging from the near-infrared to the visible spectrum. Under high pressure, the simple hexagonal (*sh*) phase of Mg is also an electride, and exhibits a hyperbolic window extending as wide as 7.40 eV, exceeding the previously reported 3.1 eV of $Bi_2Te_3$[13]. Moreover, we performed comprehensive first-principles calculations on a representative set of experimentally synthesized electrides, all of which are found to exhibit pronounced hyperbolic dispersion. We also predict the unique coexisting presence of in-plane hyperbolic plasmons and hyperbolic exciton-polaritons in some two-dimensional (2D) electrides. Unlike hyperbolic phonon-polaritons, which are restricted to a narrow window in the mid- to far-infrared range, the hyperbolic dispersion in electrides encompasses a broad spectral range, from the far-infrared to the far-ultraviolet. These discoveries highlight the unique optical properties of electrides and suggest their potential for applications in photonics.

## Results

**Limitations of Hyperbolic Dispersion in General Anisotropic Crystals**

We first present the reasoning why general anisotropic crystals cannot guarantee a hyperbolic dispersion. In metals, the hyperbolic plasmonic modes arise from the interaction between intraband and interband transitions. Intraband transitions contribute a negative real part to the dielectric function ($\varepsilon_1$), whereas interband transitions may lead to a positive $\varepsilon_1$. The interplay between these two types of transitions can result in a sign change in $\varepsilon_1$[35,36]. However, this competition typically takes effect only when near the plasma frequency ($\omega_p$). Below $\omega_p$, strong screening from free electrons makes the intraband transitions dominant. In this regime, even though interband transitions may still contribute, it is unlikely to make $\varepsilon_1$ become positive; usually it exhibits an oscillatory behavior instead. This suggests that for a metal to show hyperbolic dispersion, its $\omega_p$ must vary significantly along different crystallographic directions—a principle clearly demonstrated in an earlier study on monolayer $MoOCl_2$[37].

The tensor form of $\omega_p$ can be expressed as[38]:

$$\bar{\omega}_{p_{\alpha\beta}}^2 = \frac{4\pi e^2}{V} \sum_{n\mathbf{k}} 2 f_{n\mathbf{k}} \frac{\partial^2 \varepsilon_{n\mathbf{k}}}{\partial \mathbf{k}_\alpha \partial \mathbf{k}_\beta} \quad (0.1)$$

Because the free electrons are uniformly distributed in metals, the curvature of the energy bands with respect to momenta ($\frac{\partial^2 \varepsilon_{n\mathbf{k}}}{\partial \mathbf{k}_\alpha \partial \mathbf{k}_\beta}$) in different crystallographic directions tends to be very similar. Furthermore, summing over all eigenstates and integrating over the entire Brillouin zone introduces an averaging and smearing effect that reduces the difference in $\omega_p$ along different directions for most anisotropic metals. On the other hand, when electron scattering effects are neglected, the real part $\varepsilon_1$ arising from intraband transitions can be expressed as:

$$\varepsilon_1 = 1 - \frac{\omega_p^2}{\omega^2}$$

(0.2)

which shows a pronounced anisotropy at low frequencies for anisotropic metals, but the anisotropy diminishes rapidly as the frequency increases. Given the large value of $\omega_p$ in most metals at a scale of 10eV, the behavior of their dielectric function typically resembles that depicted in Fig. 1(a). Common structurally anisotropic metals such as Ca, Sn, and Cd all conform to this trend[39,40], and thus are difficult to achieve a hyperbolic dispersion.

Unlike the Drude-like behavior of intraband transitions in metals, the dielectric response of materials dominated by interband transitions, excitonic absorption, or phonon-polariton absorptions follow Lorentzian resonances. The real part of their dielectric function ($\varepsilon_1$) can change its sign only when near the respective resonance absorption frequencies[13]. For phonon-polariton absorption, the dielectric anisotropy is primarily due to the anisotropy in atomic positions in the crystal, there exists different roots or zero-point frequencies of $\varepsilon_1$ for different directions. However, the band gap and exciton binding energy of (anisotropic) semiconductors are quite similar along different directions, although some directional dependence exists in some electron-dominated dielectric function (such as the noticeable differences in the peak absorption of the imaginary part ($\varepsilon_2$) of the dielectric function along various directions). Consequently, the frequency at which the $\varepsilon_1$ crosses zero remains nearly identical for different directions [Fig. 1(b)].

**Unique Electronic Anisotropy in Non-Cubic Electrides**

It is evident that in order to overcome this difficulty, a completely new mechanism is required; And electrides provides such an opportunity. In particular, for electrides with non-cubic symmetry, the condition for above reasoning may not hold. Electrides are characterized by non-nuclear attractors and the subsequent nucleus-charge separation, therefore naturally bear local polarization and electronic anisotropy. Such a degree of freedom is absent in normal materials. In momentum space, the anisotropy of electrides arises from extremely narrow—often nearly flat—bands near the Fermi level, resulting from the localization of excessive charge at interstitial sites[41]. The effects of the flattened bands are triple: (i) making the local shape of the band near the energy minima or maxima away from a symmetric elliptical shape, and therefore the Fermi surface usually becomes anisotropic; (ii) the band energy variation along different directions for flat bands is hardly similar, since its value is small and becomes very sensitive to the subtle local feature; (iii) flat bands lead to heavy effective mass of electrons[28], which will shift down $\omega_p$ greatly. The former two enhance the degree of anisotropy, while the last one reduces the impact of Eq. (1.2) to $\varepsilon_1$. Finally, because many excessive charges are localized to form ISQ in electrides, the mobile carriers are significantly reduced compared to normal metals, which again reduces the value of $\omega_p$ further, and weakens the screening of the system. All of these factors sum up to strong dielectric anisotropy and weak Drude-like behavior, and thus favor hyperbolic dispersion in non-cubic electrides. However, certain exceptional cases may exist where charge localization is insufficient and ISQ states do not dominate the bands near the Fermi level. In such scenarios, even electrides with non-cubic symmetry might not exhibit hyperbolic dispersion.

It is necessary to point out that moving forward from above qualitative analysis and formulating a quantitative theory about hyperbolic dispersion in electrides is very difficult, since it heavily relies on computationally demanding numerical calculations. For the sake of clarity, in this work we will demonstrate this general feature of non-cubic electrides by a thorough screening of typical electrides, rather than develop a formal mathematical theory.

**Hyperbolic dispersion in elemental electrides**

We begin with a group of elemental crystals: magnesium (Mg), beryllium (Be) and sodium (Na). Mg exists in a non-electride phase with a hexagonal close-packed (*hcp*) structure at ambient conditions [Fig. 2(a)]. Its dielectric function exhibits typical metallic characteristics [Fig. 2(b)]: At frequencies close to zero, its $\varepsilon_1$ shows a large negative value and strong anisotropy; As frequency increases, the $\varepsilon_1$ gradually approaches zero, with the anisotropy diminishes and eventually becomes nearly isotropic at around 4 eV. Due to the interband transitions, its $\varepsilon_1$ shows a notable oscillation at around 0.75 eV. However, since this frequency is far away from the plasma frequency (at 10.63 eV), the strong intraband contributions prevent this oscillation from inducing a sign change in $\varepsilon_1$. Our calculations precisely reproduce this feature and align well with the experimental data[42,43].

When compressing Mg to a pressure of 800 GPa and entering the electride phase, the simple hexagonal (*sh*) structure of Mg becomes the stable lattice structure, and the phase is characterized by the significant charge localization into interstitial sites and the anisotropic distribution of ISQs along (*a,b*) and *c* directions[44]. Accordingly, its dielectric function also displays a significant anisotropy, as shown in the upper panel of Fig. 2(c). The decrease in $\omega_p$ along the (*a,b*) direction is small, and within the frequency range displayed here, the $\varepsilon_1$ remains negative, resembling the overall trend of gold[45]. In contrast, the $\omega_p$ along the *c* direction splits and decreases drastically to 9.76 eV and 2.36 eV, respectively, due to strong interband absorption and weakened intraband contribution. As a result, the $\varepsilon_1$ along c direction changes its sign from negative to positive at 9.76 eV, and then changes back to the negative sign at 2.36 eV, whereas the $\varepsilon_1$ along (*a,b*) direction maintains the negative value within the whole range, making $\varepsilon_1$ possesses a different sign along *c* and (*a,b*) directions, thus displaying hyperbolic dispersion behavior. The width of this hyperbolic window is exceptionally large, with a value of 7.4 eV [shaded area in Fig. 2(c)], exceeding the 3.1 eV reported for $Bi_2Te_3$[13]. In order to verify that this awesome hyperbolic behavior is indeed originated from localized states in electride rather than the lattice crystalline symmetry, we release the *sh* Mg to 0 GPa while maintaining the lattice structure. The only difference from the high pressure one is that this low pressure *sh* Mg phase ceases to be an electride [Fig.2(a)], with all excessive charges restoring to bind to nucleus again. As shown in the lower panel in Fig.2(c), its dielectric function also goes back to that of a normal metal, with a typical Drude-like $\varepsilon_1$ within the 0-7 eV frequency range, and the anisotropy disappears almost completely. This unambiguously proves that the electride plays the essential role.

Similar to Mg, Be also adopts an *hcp* structure under ambient conditions. However, it is a typical electride, as confirmed by our ELF calculation and charge density analysis (see Fig.S3). The Wyckoff sites of the two ISQs are 2*d*, which are symmetric to the two Be atoms located at 2*c*. The Bader charge is 0.61 $e^-$ for each ISQ. Unlike the *hcp* Mg that is not electride, our calculations show that for this electride phase of Be, the plasma frequencies in the (*a,b*) and *c* directions are as low as 1.94 and 1.06 eV, respectively, and acquiring a hyperbolic window of 0.88 eV between the near-infrared and visible light. This width, however, is larger than that of most known natural hyperbolic materials[12]. It is interesting to note that Weaver *et al*. had measured the dielectric function of single-crystal Be in 1973[46], and gave exactly the same characteristics, but were short of recognizing it as an NHM. Their reported dielectric function aligns closely with ours, with the plasma frequencies along the (*a,b*) directions are slightly smaller than our results. Assuming that the crystalline grains in polycrystalline material are randomly oriented, the plasma frequency of polycrystalline Be can be represented as $\omega_p = (2\omega_{p_{a,b}} + \omega_{p_c})/3$, which gives $\omega_p$ = 1.65 eV according to our calculated results. This value agrees very well with the experimental data for polycrystalline Be films of 1.6 eV[47]. Such a low $\omega_p$ is very rare among typical metals, especially if considering the fact that the carrier concentration in Be is higher than that of most metallic elements[48]. Based on the classical Drude model: $\omega_p = \sqrt{\frac{ne^2}{\varepsilon_0 m^*}}$, where *n* is the carrier concentration and *m*\* is the effective mass of electron, $\omega_p$ decreases if *m*\* becomes large. We found that *m*\* in Be is indeed very large, due to the influence of localized ISQs. The projected band structure of *hcp* Be is shown in Fig. 3(a). It illustrates that the bands near the Fermi level are dominated by ISQs, and are flat. Which give rise to a small band curvature, and thus a large effective mass for the carriers.

With the frequency increases, the sign of $\varepsilon_1$ changes again, due to the interband transitions, a mechanism that is similar to the elements in the P electron block[49]. As a result, $\varepsilon_1$ changes from positive to negative at a frequency of 3.92 eV along the ($a,b$) direction and 4.40 eV along the $c$ direction, making Be have a second hyperbolic window [see Fig. 2(d)]. It should be pointed out that imaginary part ($\varepsilon_2$) does not exhibit significant absorption peak at the corresponding frequency. This is because the energy states near the Fermi level are continuous in metallic systems, such that the electrons are allowed to move freely within a finite energy range. For this reason, the interband transitions in metals typically do not form clear absorption peaks, but instead manifest as a relatively broad absorption band.

Compared to the prototypical layered electrides $Ca_2N$ and $Sr_2N$[33,34], *hcp* Be and high-pressure *sh* Mg do not exhibit a particularly low $\varepsilon_2$. This difference stems from the fact that in $Ca_2N$ and $Sr_2N$, the interstitial electron states form bands that lie right at the Fermi level and are well separated from other bands. Such a band structure strongly suppresses photoinduced electronic transitions, keeping $\varepsilon_2$ at an exceptionally low level and thereby minimizing dielectric losses.

In contrast, *hcp* Be and high-pressure *sh* Mg feature interstitial states that are less cleanly separated from the background bands, $\varepsilon_2$ suppression is less effective. Nevertheless, within their hyperbolic frequency windows $\varepsilon_2$ still declines markedly and shows no pronounced peaks. Compared with conventional hyperbolic metamaterials—which rely on metals' high losses below the plasma frequency to achieve a negative real permittivity $\varepsilon_1$—*hcp* Be and *sh* Mg under pressure maintain dielectric losses that remain acceptable for practical applications.

Another notable elemental electride is *hp4* Na, which exists at high pressures above 200 GPa and is an insulating electride[50]. The ISQ positions are similar to those of *hcp* Be, but with stronger localization. Figure 3(b) presents the projected band structure and DOS for *hp4* Na at 230 GPa, as calculated by using the $G_0W_0$ approximation. It has a band gap of 1.40 eV; and the valence bands near the Fermi level are primarily governed by ISQ states, which show unusually flat band features. The dielectric function, computed by using the BSE method, is shown in Fig. 2(e). From $\varepsilon_2$, it is evident that the absorption edge along the *c* direction is close to the bandgap, while along the (*a,b*) directions, the absorption edge exceeds the bandgap by about 1.10 eV. That is, this material is transparent along (*a,b*) direction, but becomes opaque when along the *c* direction for visible light, a typical orientation-dependent transparent-opaque dual material. Additionally, the main absorption peaks along the (*a,b*) and *c* directions are shifted apart by about 1.93 eV. This substantial difference would typically produce a broad hyperbolic window. However, the absorption peak along the *c* direction exhibits split double peaks, causing the real part in this direction to briefly become negative near the first peak before quickly returning to positive values. It becomes consistently negative only after the second peak, leading to a reduced hyperbolic window between 3.56 eV and 3.81 eV. The wide hyperbolic dispersion in *hp4* Na driven by anisotropic interband transitions is quite rare; even for layered materials like $Bi_2Te_3$, the hyperbolic window in both in-plane and out-of-plane directions, as induced by interband transitions, does not exceed 0.1 eV[51].

**Hyperbolic dispersion in binary and ternary non-cubic electrides**

The aforementioned elemental electrides, except for Be, are presented at high pressures. Nonetheless, materials accessible at ambient conditions are essential for practical applications. Therefore, we also explored the binary electrides that have been successfully synthesized in ambient-pressure experiments to consolidate the above argument; and some typical examples of them are demonstrated below. TiH is one of the stable products formed during the hydrogenation of face-centered cubic Ti, with a space group of *P4$_2$/mmc*[52]. As illustrated by the ELF in Fig. 4(a), the valence electrons of Ti are transferred to H atom or localized into the lattice interstitial sites and forming electride phase. The overall trend of the real part of the dielectric function of TiH is similar to that of Be, with a lower $\omega_p$ of 0.32 eV along the (*a,b*) directions, and a higher $\omega_p$ of 1.84 eV along the *c* direction [Fig. 4(d)]. Due to the strong anisotropy in interband transitions along different directions, a second hyperbolic window also emerges between 2.22 and 3.09 eV. This indicates that the wide hyperbolic window of TiH spans from near-infrared to the entire visible spectrum. Additionally, we also check the dielectric function by ignoring the local field effects [Fig. 4(d)], which gives almost identical results with those include local field effects. The same conclusion is obtained for all metallic systems in this study, which suggests that forming ISQs in electrides does not introduce significant local field effects, and also confirming the reliability of our calculated dielectric function.

$Sc_2Sb$ and $Be_2Zr$ are also binary electrides that have been successfully synthesized experimentally, belonging to a space group *P4/nmm* and *P6/mmm*, respectively[53,54]. Like TiH, they exhibit prominent interstitial charge localization [Fig. 4(b,c)]. $Sc_2Sb$ is weak metallic. Under the influence of ISQs, its plasma frequencies are shifted down significantly, and in the (*a,b*) and *c* directions are only 0.13 eV and 0.98 eV, respectively, corresponding to its first hyperbolic window. Two additional hyperbolic windows, driven by anisotropic interband transitions, appear at frequencies of 1.47-1.94 eV and 3.04-3.28 eV, respectively.

In contrast, $Be_2Zr$ exhibits slightly stronger metallic character than TiH and $Sc_2Sb$ (Their band structures and DOS are shown in Fig. S1 of the Supplementary Information). It has an $\omega_p$ of 2.30 eV and 3.53 eV along the (*a,b*) and *c* directions, respectively. Similar to *hcp* Mg, the influence of interband transitions also causes fluctuations in the $\varepsilon_1$ of $Be_2Zr$ at around 1 eV; but it is not strong enough to change the sign. Compared with *hcp* Be, these three binary bulk electrides exhibit a slightly reduced imaginary part, indicating lower dielectric losses in practical applications.

It should be noted that in most of the aforementioned examples, ISQs exhibit zero-dimensional (0D) localization. However, from a theoretical standpoint, non-cubic electrides with any type of ISQ are highly likely to exhibit hyperbolic dispersion. In other words, the specific spatial topology, dimensionality, and local symmetry of the ISQs may influence certain details—such as the width of the hyperbolic window and the magnitude of dielectric loss—but do not determine whether hyperbolic dispersion occurs.

Taking the ternary electride $Ba_3LiN$ as an example[55,56], its interstitial electrons possess more spatial degrees of freedom, and the ISQs exhibit a three-dimensional (3D) localized character that differs from the previous examples [see Fig. 5 (a)]. Nevertheless, this material still displays pronounced hyperbolic dispersion. As shown in Fig. 5 (b), the plasma frequency along the c-axis is as low as 1.00 eV, while the corresponding values along the a and b axes are 4.62 eV. Although the real part of the dielectric function along the c-axis also shows oscillatory behavior around zero beyond the plasma frequency—similar to the previously discussed materials—it remains opposite in sign to the in-plane components within the energy range of 1.00–2.67 eV, thereby exhibiting a typical hyperbolic dispersion.

**In-plane hyperbolic dispersion in two-dimensional electrides**

Clearly, the general rule for the hyperbolic dispersion and plasmon frequency reduction induced by ISQs also applies to electrides with 2D lattice structure (termed as 2D electride below for brevity). Namely, as long as the structural anisotropy in the *a* and *b* directions of 2D electrides presents, ISQs will amplify such anisotropy, and an in-plane hyperbolic dispersion can be achieved. Here, we demonstrate $Sc_5Cl_8$, $Y_2Cl_3$, and $Ca_2Cu$ as examples. They all are experimentally already synthesized layered materials, in which $Sc_5Cl_8$ and $Y_2Cl_3$ belong to the *C2/m* space group and were proposed as candidates for exfoliation into 2D electrides[25,57]. On the other hand, $Ca_2Cu$ is in the *Pnma* space group[58], and our ELF calculations indicate that it is also a promising electride. As shown in Fig.6(a~c), the calculated ISQs in these three materials are all located in the plane, not at between the interlayer as in $Ca_2N$[34], which ensures that ISQs persist after being exfoliated to monolayers.

Monolayer $Sc_5Cl_8$ and $Ca_2Cu$ are metallic, their calculated dielectric function are shown in Fig. 6(d, e). Similar to bulk metallic electrides, the in-plane ISQs cause strong anisotropy in $\omega_p$, leading to in-plane hyperbolic plasmons. Furthermore, the very low carrier concentration in 2D systems and the high effective electron mass caused by ISQs make their $\omega_p$ locate in the infrared region. Such a small $\omega_p$ implies that their hyperbolic window range cannot be as large as that of bulk electrides. Nevertheless, the in-plane hyperbolic windows of monolayer $Sc_5Cl_8$ and $Ca_2Cu$ are still 4-5 times larger than that of $WTe_2$, with a hyperbolic window of 0.48-0.60 eV for $Sc_5Cl_8$ and 0.11-0.21 eV for $Ca_2Cu$, respectively, compared to 0.053-0.078 eV in $WTe_2$[19]. It should also be noted that $Ca_2Cu$ has a second hyperbolic window from 0.32~0.41 eV, making it a unique 2D system that host two separate hyperbolic windows that reciprocal to each other in the orientational direction of the dispersion hyperbola [Fig.5(e)].

Monolayer $Y_2Cl_3$ is a 2D semiconducting electride. Its dielectric function is shown in Fig. 6(f). It has one excitonic absorption peak for each direction, at 1.46 eV (*a* direction) and 1.62 eV (*b* direction), respectively. This difference in the exciton peak energy of two directions suggests significant anisotropy in this electrides. Most importantly, like *hp4* Na, the hyperbolic dispersion in $Y_2Cl_3$ is also induced by interband transitions, with the interband absorption edge along the *b* direction at 1.90 eV (close to its bandgap of 1.91 eV), and the interband absorption edge along the *a* direction at 1.6 eV. The interband absorption peaks occur at 2.39 (*a* direction) and 1.96 (*b* direction) eV, respectively, resulting in further decline of $\varepsilon_1$ in each direction. Specifically, along the *b* direction, the $\varepsilon_1$ decreases significantly at 2.27 eV and then crosses zero; while $\varepsilon_1$ along *a* direction always maintains positive, creating a hyperbolic window from 2.41 to 2.60 eV. This hyperbolic window surpasses those of the vast majority of known in-plane hyperbolic materials and is especially valuable because it falls within the visible spectrum. Moreover, within this window, monolayer $Y_2Cl_3$ exhibits an imaginary part $\varepsilon_2$ as low as that of monolayer $Sc_5Cl_8$, and as an intrinsic 2D in-plane hyperbolic material, it inherently circumvents the additional losses introduced by finite thickness or patterning. Finally, it should be pointed out that though the excitonic peaks in $Y_2Cl_3$ at 1.46 eV do not cause $\varepsilon_1$ to turn negative and form hyperbolic exciton-polaritons, it is anticipated that in other anisotropic 2D electrides with more intense excitonic absorptions, hyperbolic exciton-polaritons may be easy to observe.

## Conclusions

In summary, we elucidated and demonstrated that ISQs in non-cubic electrides have profound impacts on the optical response: it can reduce $\omega_p$ into visible and infrared light range, bring strong anisotropy and multiple absorption peaks, and makes $\varepsilon_1$ change sign at different frequencies. All of these make non-cubic electrides promising candidates for NHMs, exhibiting diverse dispersion characteristics and broad hyperbolic windows. In particular, we showed that the elemental metal Be under ambient conditions has two hyperbolic windows in the visible and ultraviolet light regions; high-pressure *sh* Mg has an exceptionally wide hyperbolic window from 2.36 eV to 9.76 eV. In semiconducting electrides, the hyperbolic dispersion can be simply induced by anisotropic interband absorption. Based on this mechanism, 2D $Y_2Cl_3$ exhibits an ultra-wide in-plane hyperbolic dispersion window that lies within the visible light range. These outstanding properties originate from the localized charge distribution and the resulting extremely narrow energy bands. It is interesting to note that strong correlation system such as lanthanides or actinides that contain localized f electrons also exhibits flat bands. They also could induce hyperbolic dispersion. However, electride is better on this issue because: (i) localized f electrons are tightly bound to nucleus, thus has little freedom to amplify the lattice anisotropy by comparing with electride, in which ISQ has great space to deform to magnify local polarization; (ii) localized f orbitals are usually away from the Fermi-level, and immersed in the electron sea of *s*/*p* states, Thus the anisotropy is weakened by the screening of these isotropic valence electrons. Electride is unique since ISQs are always on or exposed to Fermi surface. The general understanding we obtained in this work, as well as the diverse and excellent hyperbolic properties of electrides presented here, is expected to influence the current strategy for searching and designing NHMs, potentially shifting the focus from layered or highly anisotropic materials to non-cubic electrides. The latter seems to offer more promising possibilities.

## Methods

Our first-principles calculations start with density functional theory (DFT) with the Perdew-Burke-Ernzerhof (PBE)[59] exchange-correlation functional and the projector augmented-wave potentials (PAW)[60,61] to describe the electron-ion interactions. Based on the obtained self-consistent DFT wavefunctions, a single-shot GW (i.e., $G_0W_0$)[62-64] calculation is performed, alongside calculations of the dielectric function of metals using linear response theory[38,65,66]. The local field effects are included at the Hartree level. The energy cutoff for the response function in the $G_0W_0$ calculations is set to 310 eV or higher, while the cutoff energy for plane-wave basis is set to 1.5 to 2 times of this value. Since the dielectric function of metals at GW approximation is challenging to converge with respect to k-points, especially in the low-frequency range, we use the largest feasible k-point grids to ensure high accuracy: for some metallic systems, the Monkhorst-Pack[67] k-points grid spacing is as fine as $2\pi \times 0.007$ Å$^{-1}$. Furthermore, for semiconductors, we perform additional Bethe-Salpeter equation (BSE)[68] calculations on top of the $G_0W_0$ wave-functions to obtain a dielectric function that taking excitonic effects into account. It should be noted that the dielectric function of the 2D systems obtained through calculations is more qualitative, as its numerical values may be influenced by the vacuum layer thickness used in the calculation[69]. However, the vacuum layer thickness does not significantly affect the trend of the dielectric function with frequency, making it sufficient for describing hyperbolic dispersion. All calculations are conducted using the VASP package.[70,71] The structural parameters used for electronic structure and optical property calculations are provided in Table S1 of the Supplementary Information.

## Data availability

The data that support the findings of this study are available from the corresponding author upon reasonable request.

## Competing interests

The authors declare no competing interests.

# Acknowledgments

This work was supported by National Key R&D Program of China under Grant No. 2021YFB3802300, the NSAF under Grant Nos. U1730248, the National Natural Science Foundation of China under Grant No. 12074274, and the Foundation of National key Laboratory of shock wave and Detonation physics under Grant No. 2023JCJQLB05401. Part of the calculations were performed using the resources provided by the center for Comput. Mater. Sci. (CCMS) at Tohoku University, Japan.

# Author Contribution

Qi-Dong. Hao: Calculation, Analysis, Writing, Original draft preparation. Hao. Wang: Analysis, Validation. Hong-Xing. Song: Analysis, Validation. Xiang-Rong. Chen: Writing, Reviewing and Editing, Fund. Hua-Yun. Geng: Idea conceiving, Project design, Writing, Reviewing and Editing, Fund.

# References

1   Yao, J., Yang, X., Yin, X., Bartal, G. & Zhang, X. Three-dimensional nanometer-scale optical cavities of indefinite medium. *Proc. Natl. Acad. Sci. U. S. A.* **108**, 11327-11331 (2011).
2   Drachev, V. P., Podolskiy, V. A. & Kildishev, A. V. Hyperbolic metamaterials: new physics behind a classical problem. *Opt. Express* **21**, 15048-15064 (2013).
3   Lu, W. & Sridhar, S. Superlens imaging theory for anisotropic nanostructured metamaterials with broadband all-angle negative refraction. *Phys. Rev. B* **77**, 233101 (2008).
4   Liu, Z., Lee, H., Xiong, Y., Sun, C. & Zhang, X. Far-field optical hyperlens magnifying sub-diffraction-limited objects. *Science* **315**, 1686-1686 (2007).
5   Shelby, R. A., Smith, D. R. & Schultz, S. Experimental verification of a negative index of refraction. *Science* **292**, 77-79 (2001).
6   Smith, D. & Schurig, D. Electromagnetic wave propagation in media with indefinite permittivity and permeability tensors. *Phys. Rev. Lett.* **90**, 077405 (2003).
7   Poddubny, A., Iorsh, I., Belov, P. & Kivshar, Y. Hyperbolic metamaterials. *Nat. Photonics* **7**, 948-957 (2013).
8   Dai, S. *et al.* Subdiffractional focusing and guiding of polaritonic rays in a natural hyperbolic material. *Nat. Commun.* **6**, 6963 (2015).
9   Korzeb, K., Gajc, M. & Pawlak, D. A. Compendium of natural hyperbolic materials. *Opt. Express* **23**, 25406-25424 (2015).
10  Gjerding, M. N., Petersen, R., Pedersen, T. G., Mortensen, N. A. & Thygesen, K. S. Layered van der Waals crystals with hyperbolic light dispersion. *Nat.*

11   Dai, S. *et al.* Tunable phonon polaritons in atomically thin van der Waals crystals of boron nitride. *Science* **343**, 1125-1129 (2014).

12   Sun, J., Litchinitser, N. M. & Zhou, J. Indefinite by nature: from ultraviolet to terahertz. *ACS Photonics* **1**, 293-303 (2014).

13   Esslinger, M. *et al.* Tetradymites as natural hyperbolic materials for the near-infrared to visible. *ACS Photonics* **1**, 1285-1289 (2014).

14   Sun, J., Zhou, J., Li, B. & Kang, F. Indefinite permittivity and negative refraction in natural material: Graphite. *Appl. Phys. Lett.* **98**, 101901 (2011).

15   Ma, W. *et al.* In-plane anisotropic and ultra-low-loss polaritons in a natural van der Waals crystal. *Nature* **562**, 557-562 (2018).

16   Zheng, Z. *et al.* A mid-infrared biaxial hyperbolic van der Waals crystal. *Sci. Adv.* **5**, aav8690 (2019).

17   Wang, R., Sun, J. & Zhou, J. Indefinite permittivity in uniaxial single crystal at infrared frequency. *Appl. Phys. Lett.* **97** (2010).

18   Taboada-Gutiérrez, J. *et al.* Broad spectral tuning of ultra-low-loss polaritons in a van der Waals crystal by intercalation. *Nat. Mater.* **19**, 964-968 (2020).

19   Wang, C. *et al.* Van der Waals thin films of $WTe_2$ for natural hyperbolic plasmonic surfaces. *Nat. Commun.* **11**, 1158 (2020).

20   Ruta, F. L. *et al.* Hyperbolic exciton polaritons in a van der Waals magnet. *Nat. Commun.* **14**, 8261 (2023).

21   Wang, F. *et al.* Prediction of hyperbolic exciton-polaritons in monolayer black phosphorus. *Nat. Commun.* **12**, 5628 (2021).

22   Miao, M. S. & Hoffmann, R. High pressure electrides: a predictive chemical and physical theory. *Acc. Chem. Res.* **47**, 1311-1317 (2014).

23   Hosono, H. & Kitano, M. Advances in materials and applications of inorganic electrides. *Chem. Rev.* **121**, 3121-3185 (2021).

24   Zhang, Y., Wang, H., Wang, Y., Zhang, L. & Ma, Y. Computer-assisted inverse design of inorganic electrides. *Phys. Rev. X* **7**, 011017 (2017).

25   Zhou, J. *et al.* Discovery of hidden classes of layered electrides by extensive high-throughput material screening. *Chem. Mater.* **31**, 1860-1868 (2019).

26   Zhang, X. *et al.* Magnetic electrides: High-throughput material screening, intriguing properties, and applications. *J. Am. Chem. Soc.* **145**, 5523-5535 (2023).

27   Hao, Q. D., Wang, H., Chen, X. R. & Geng, H. Y. Distinctive electronic characteristics and ultra-high thermoelectric power factor of Be–Fe intermetallics. *J. Mater. Chem. A,* **12**, 24370-24379 (2024).

28   Zhang, L., Geng, H. Y. & Wu, Q. Prediction of anomalous LA-TA splitting in electrides. *Matter Radiat. Extremes* **6**, 038403 (2021).

29   Zhang, L. *et al.* Interplay of anionic quasi-atoms and interstitial point defects in electrides: abnormal interstice occupation and colossal charge state of point defects in dense fcc-lithium. *ACS Appl. Mater. Interfaces* **13**, 6130-6139 (2021).

30   Wang, D. *et al.* Universal Metallic Surface States in Electrides. *J. Phys. Chem. C* **128**, 1845-1854 (2024).

31. Wang, D. et al. Simultaneous Superconducting and Topological Properties in Mg–Li Electrides at High Pressures. *J. Phys. Chem. C* **129**, 689-698 (2024).

32. Yu, Z., Geng, H. Y., Sun, Y. & Chen, Y. Optical properties of dense lithium in electride phases by first-principles calculations. *Sci. Rep.* **8**, 3868 (2018).

33. Guan, S., Yang, S. A., Zhu, L., Hu, J. & Yao, Y. Electronic, dielectric and plasmonic properties of two-dimensional electride materials $X_2N$ (X= Ca, Sr): a first-principles study. *Sci. Rep.* **5**, 12285 (2015).

34. Guan, S., Huang, S. Y., Yao, Y. & Yang, S. A. Tunable hyperbolic dispersion and negative refraction in natural electride materials. *Phys. Rev. B* **95**, 165436 (2017).

35. Nemilentsau, A., Low, T. & Hanson, G. Anisotropic 2D materials for tunable hyperbolic plasmonics. *Phys. Rev. Lett.* **116**, 066804 (2016).

36. Wang, H. et al. Planar hyperbolic polaritons in 2D van der Waals materials. *Nat. Commun.* **15**, 69 (2024).

37. Zhao, J. et al. Highly anisotropic two-dimensional metal in monolayer $MoOCl_2$. *Phys. Rev. B* **102**, 245419 (2020).

38. Harl, J. The linear response function in density functional theory. Ph.D. thesis, University of Vienna (2008).

39. Graves, R. H. W. & Lenham, A. P. Determination of the optical constants of uniaxial or isotropic metals by measurement of reflectance ratios. *J. Opt. Soc. Am.* **58**, 884-889 (1968).

40. Graves, R. H. W. & Lenham, A. P. Interband absorption in single crystals of Mg, Zn, and Cd at 295 K and 82 K. *J. Opt. Soc. Am.* **58**, 126-129 (1968).

41. Wei, Y. H. et al. Realization of kagome lattice and superconductivity in topological electrides. *Phys. Rev. B* **110**, 054511 (2024).

42. Jones, D. & Lettington, A. H. The optical properties and electronic structure of magnesium. *Proc. Phys. Soc.* **92**, 948–955 (1967).

43. Palik, E. D. *Handbook of optical constants of solids*. Vol. 3 (Academic Press, 1998).

44. Miao, M. S. & Hoffmann, R. High-pressure electrides: the chemical nature of interstitial quasiatoms. *J. Am. Chem. Soc.* **137**, 3631-3637 (2015).

45. Rioux, D. et al. An analytic model for the dielectric function of Au, Ag, and their alloys. *Adv. Opt. Mater.* **2**, 176-182 (2014).

46. Weaver, J. H., Lynch, D. W. & Rosei, R. Optical properties of single-crystal Be from 0.12 to 4.5 eV. *Phys. Rev. B* **7**, 3537 (1973).

47. Arakawa, E. T., Callcott, T. A. & Chang, Y.-C. in *Handbook of Optical Constants of Solids* 421-433 (Elsevier, 1997).

48. Ashcroft, N. W. & Mermin, N. D. *Solid state physics*. (Saunders College Publishing, 1976).

49. Toudert, J. & Serna, R. Interband transitions in semi-metals, semiconductors, and topological insulators: a new driving force for plasmonics and nanophotonics. *Opt. Mater. Express* **7**, 2299-2325 (2017).

50. Ma, Y. et al. Transparent dense sodium. *Nature* **458**, 182-185 (2009).

51. Busselez, R., Levchuk, A., Ruello, P., Juvé, V. & Arnaud, B. Anisotropy in the

|    |    |
|----|----|
|    | dielectric function of $Bi_2Te_3$ from first principles: From the UV-visible to the infrared range. *Phys. Rev. B* **107**, 174305 (2023). |
| 52 | Numakura, H. & Koiwa, M. Hydride precipitation in titanium. *Acta Metall.* **32**, 1799-1807 (1986). |
| 53 | Nuss, J. & Jansen, M. Crystal structure of discandium antimonide, $Sc_2Sb$. *New Cryst. Struct.* **217**, 19-20 (2002). |
| 54 | Tanner, L. & Ray, R. Metallic glass formation and properties in Zr and Ti alloyed with Be—I the binary Zr-Be and Ti-Be systems. *Acta Metall.* **27**, 1727-1747 (1979). |
| 55 | Smetana, V., Babizhetskyy, V., Vajenine, G. V. & Simon, A. Synthesis and crystal structure of LiBa2N and identification of $LiBa_3N$. *J. Solid State Chem.* **180**, 1889-1893 (2007). |
| 56 | Weaver, S. M., Lanetti, M. G., Slamowitz, C. C., Radomsky, R. C. & Warren, S. C. Assessing Dimensionality in Electrides. *J. Phys. Chem. C* **129**, 3211-3217 (2025). |
| 57 | Wan, B. *et al.* Identifying quasi-2D and 1D electrides in yttrium and scandium chlorides via geometrical identification. *Npj Comput. Mater.* **4**, 77 (2018). |
| 58 | Fornasini, M. L. $Ca_2Cu$ with trigonal-prismatic coordination of the copper atoms forming infinite chains. *Acta Crystallogr., Sect. B: Struct. Crystallogr. Cryst. Chem.* **38**, 2235-2236 (1982). |
| 59 | Perdew, J. P., Burke, K. & Ernzerhof, M. Generalized gradient approximation made simple. *Phys. Rev. Lett.* **77**, 3867-3868 (1996). |
| 60 | Blöchl, P. E. Projector augmented-wave method. *Phys. Rev. B* **50**, 17953-17979 (1994). |
| 61 | Kresse, G. & Joubert, D. From ultrasoft pseudopotentials to the projector augmented-wave method. *Phys. Rev. B* **59**, 1758-1775 (1999). |
| 62 | Hedin, L. New method for calculating the one-particle Green's function with application to the electron-gas problem. *Phys. Rev.* **139**, A796 (1965). |
| 63 | Shishkin, M. & Kresse, G. Implementation and performance of the frequency-dependent GW method within the PAW framework. *Phys. Rev. B* **74**, 035101 (2006). |
| 64 | Hybertsen, M. S. & Louie, S. G. Electron correlation in semiconductors and insulators: Band gaps and quasiparticle energies. *Phys. Rev. B* **34**, 5390 (1986). |
| 65 | Gajdoš, M., Hummer, K., Kresse, G., Furthmüller, J. & Bechstedt, F. Linear optical properties in the projector-augmented wave methodology. *Phys. Rev. B* **73**, 045112 (2006). |
| 66 | Liu, P., Kaltak, M., Klimeš, J. & Kresse, G. Cubic scaling GW: Towards fast quasiparticle calculations. *Phys. Rev. B* **94**, 165109 (2016). |
| 67 | Monkhorst, H. J. & Pack, J. D. Special points for Brillouin-zone integrations. *Phys. Rev. B* **13**, 5188 (1976). |
| 68 | Onida, G., Reining, L. & Rubio, A. Electronic excitations: density-functional versus many-body Green's-function approaches. *Rev. Mod. Phys.* **74**, 601 (2002). |
| 69 | Tian, T. *et al.* Electronic polarizability as the fundamental variable in the |

dielectric properties of two-dimensional materials. *Nano Lett.* **20**, 841-851 (2019).
70  Kresse, G. & Furthmüller, J. Efficiency of ab-initio total energy calculations for metals and semiconductors using a plane-wave basis set. *Comput. Mater. Sci.* **6**, 15-50 (1996).
71  Kresse, G. & Furthmüller, J. Efficient iterative schemes for ab initio total-energy calculations using a plane-wave basis set. *Phys. Rev. B* **54**, 11169-11186 (1996).

**Figure captions**

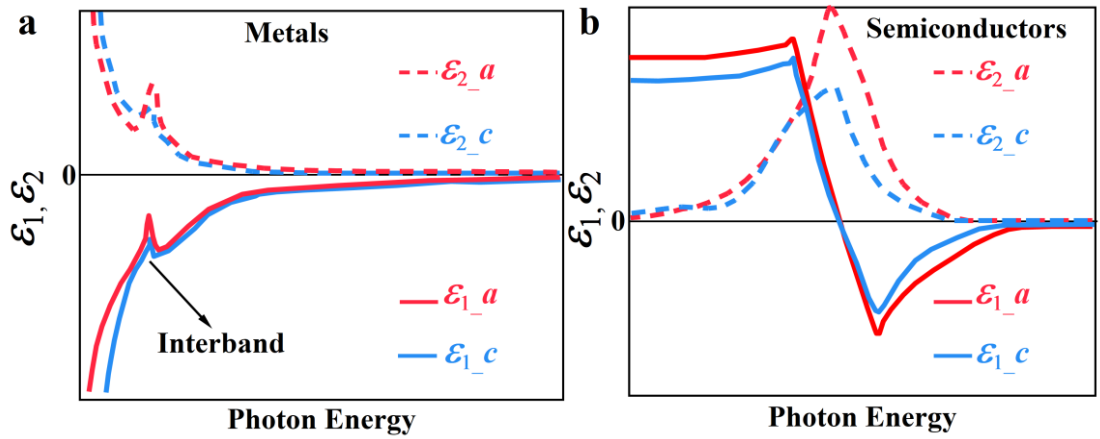

**Fig. 1: Dielectric function for general anisotropic non-layered materials.** (a) shows a metallic system; (b) shows a semiconductor system. The real part of the dielectric function ($\varepsilon_1$) is plotted as a solid line, and the imaginary part ($\varepsilon_2$) as a dashed line. Subscripts *a* and *c* denote directions perpendicular and parallel to the principal axis, respectively. In each panel, the dielectric functions along the *a* direction are shown in red (solid and dashed lines), and along the *c* direction in blue (solid and dashed lines).

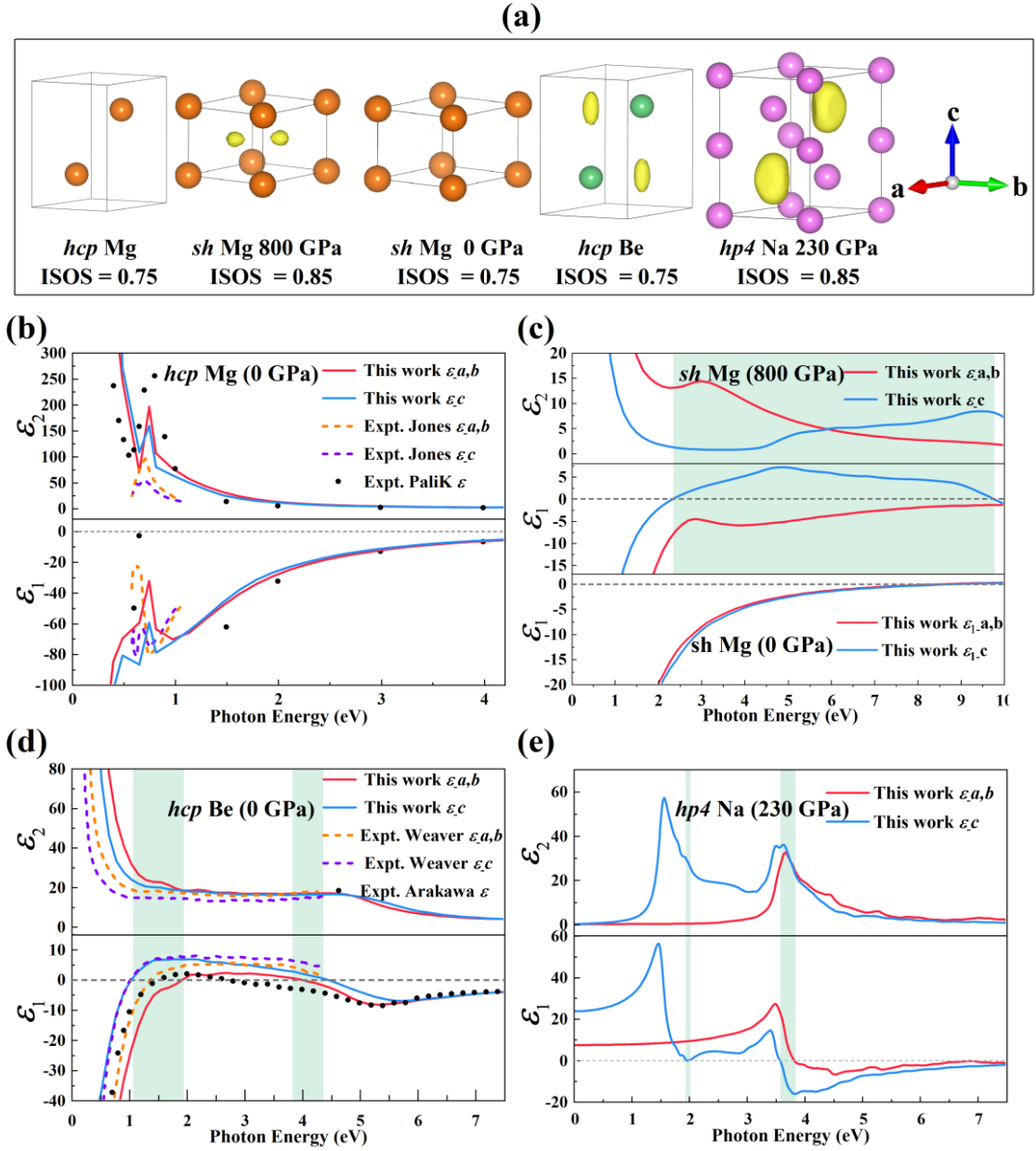

**Fig. 2: Lattice structure, electronic localization function and dielectric function for several elemental solids.** (a) The lattice structure and electronic localization function (ELF) of elemental Mg, Be, and Na, with the isosurface (ISOS) value indicated below, the yellow non-spherical area denotes ISQs. (b, c, d, e) show the dielectric function of *hcp* Mg at 0 GPa, *sh* Mg at 800 GPa and 0 GPa, *hcp* Be at 0 GPa, and *hp4* Na at 230 GPa, respectively. $\varepsilon_1$ and $\varepsilon_2$ are the real and imaginary parts of the dielectric function, respectively. Subscripts (*a,b*) and *c* denote directions perpendicular and parallel to the principal axis, respectively. In each panel, the dielectric functions along (*a,b*) direction are shown in red, and along the *c* direction in blue. Hyperbolic windows are indicated by green shaded areas. Experimental data are: dashed lines in (b) taken from Jones et al.[42], black dots

in (b) from Palik[43], dashed lines in (d) taken from Weaver et al.[46], black dots in (d) from Arakawa et al.[47], respectively.

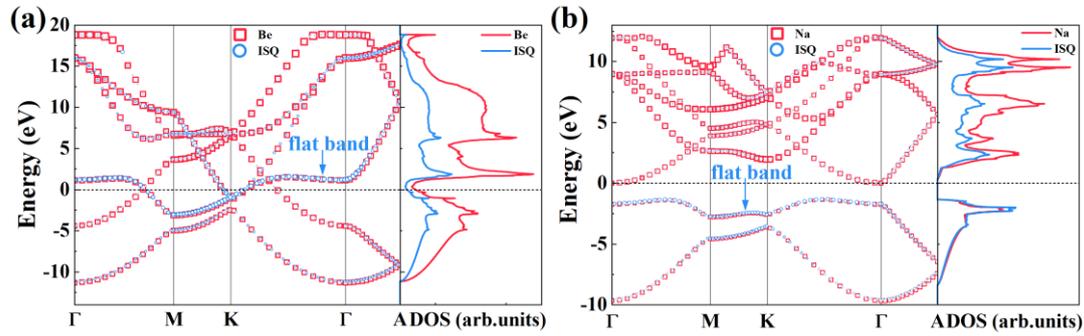

**Fig. 3: Projected band structure and density of states.** (a) *hcp* Be at 0 GPa and (b) *hp4* Na at 230 GPa. Red squares and blue circles indicate the contributions from atoms and interstitial electrons (ISQs), respectively. The main panel on the left shows the electronic bands along high-symmetry directions. The inset on the right shows the corresponding density of states (DOS), with atomic and ISQ contributions plotted as red and blue lines, respectively.

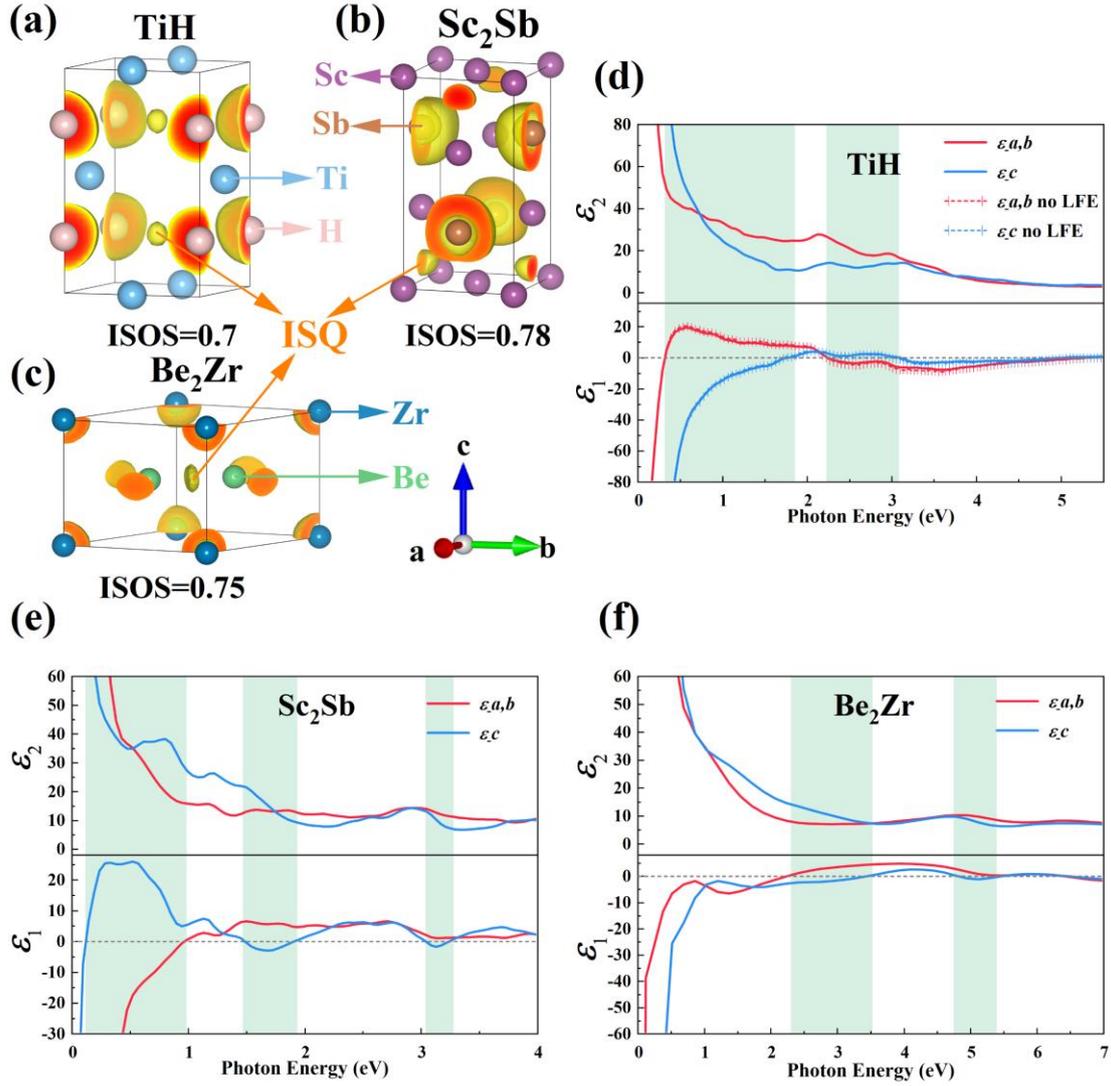

**Fig. 4: Electron localization and dielectric functions of selected binary electrides.** (a), (b), and (c) show the lattice structures and electron localization functions (ELF) of TiH, Sc$_2$Sb, and Be$_2$Zr, with the isosurface (ISOS) value indicated below, the yellow non-spherical area denotes ISQs; (d), (e), and (f) show their respective dielectric functions, the real ($\varepsilon_1$) and imaginary ($\varepsilon_2$) parts are plotted as solid and dashed lines, respectively. Subscripts (*a,b*) and *c* denote directions perpendicular and parallel to the principal axis, respectively. In each panel, the dielectric functions along (*a,b*) direction are shown in red, and along the *c* direction in blue. Hyperbolic windows are indicated by green shaded areas. $\varepsilon\_{a,b}$ no LFE and $\varepsilon\_c$ no LFE denote the dielectric functions without local field effects.

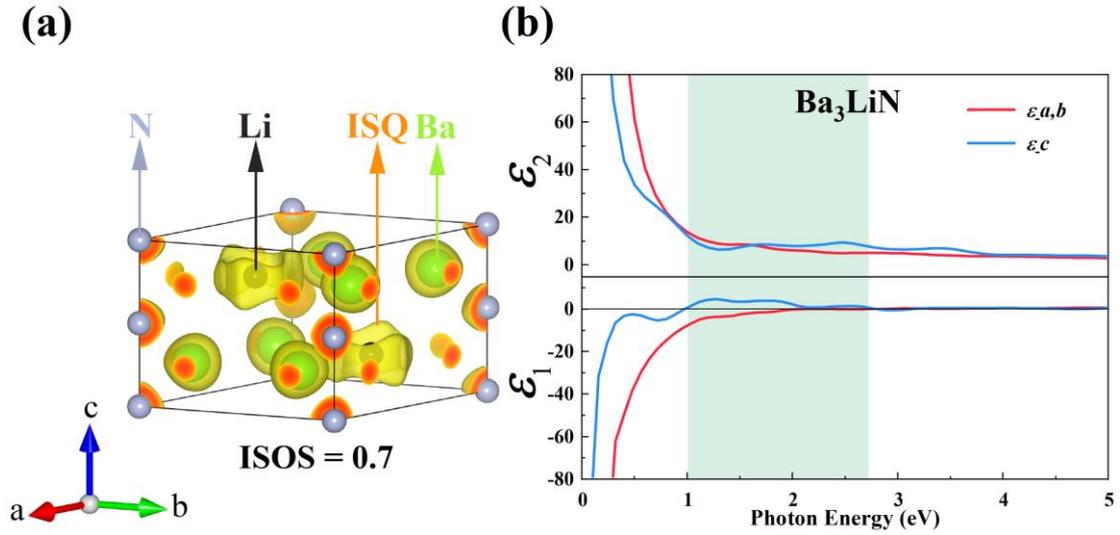

**Fig. 5: Electron localization and dielectric functions of Ba$_3$LiN.** (a) lattice structures and electron localization functions (ELF) of Ba$_3$LiN, with the isosurface (ISOS) value indicated below, the yellow non-spherical area denotes ISQs; (b) dielectric functions of Ba$_3$LiN. the real ($\varepsilon_1$) and imaginary ($\varepsilon_2$) parts are plotted as solid and dashed lines, respectively. Subscripts (*a,b*) and *c* denote directions perpendicular and parallel to the principal axis, respectively. In each panel, the dielectric functions along (*a,b*) direction are shown in red, and along the *c* direction in blue. Hyperbolic windows are indicated by green shaded areas.

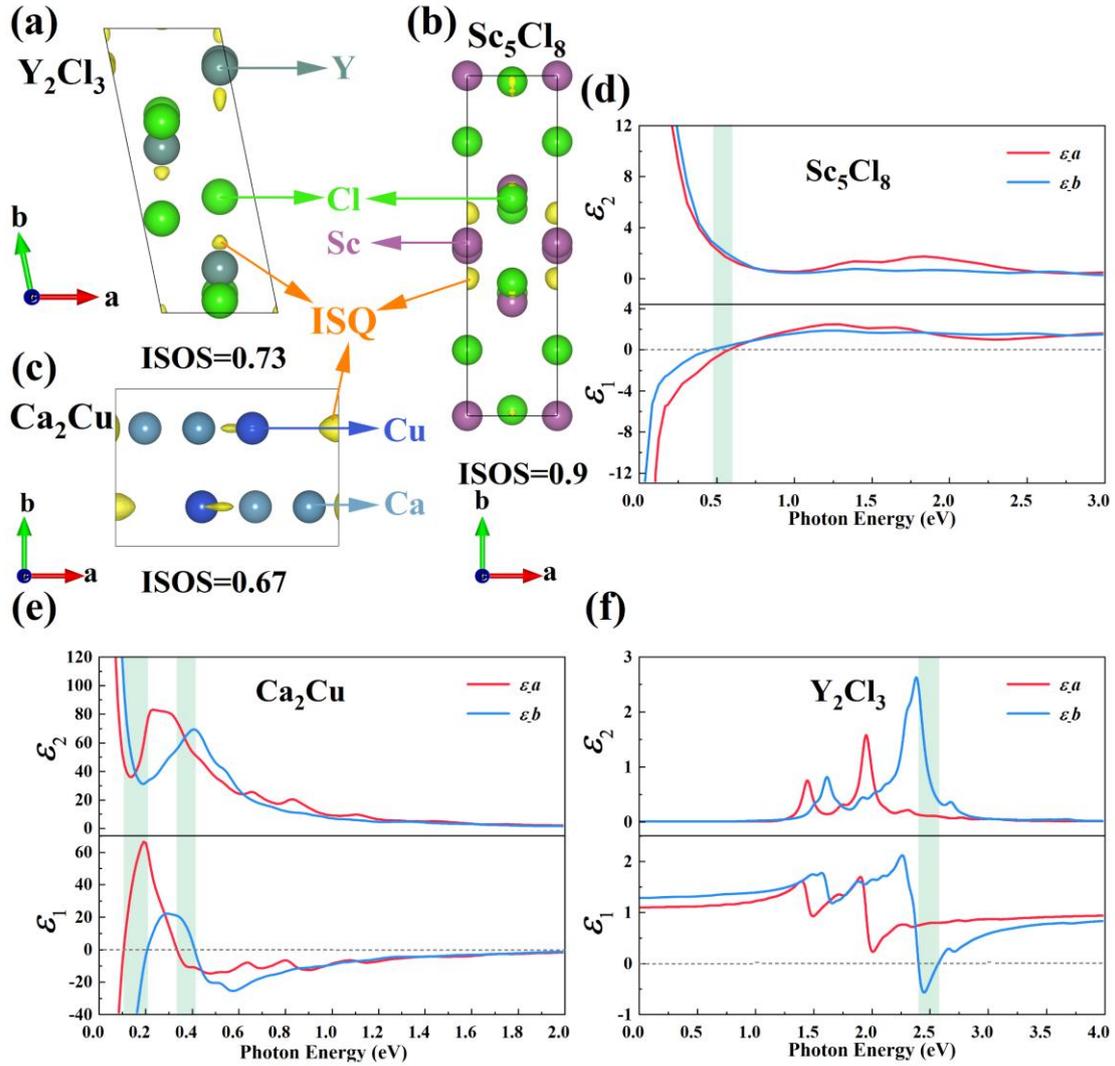

**Fig. 6: Electron localization and dielectric functions for the selected monolayer electrides.** (a), (b), and (c) show the lattice structures and electron localization functions (ELF) of $Y_2Cl_3$, $Sc_5Cl_8$, and $Ca_2Cu$, with the isosurface (ISOS) value indicated below, the yellow non-spherical area denotes ISQs; (f), (d), and (e) show their respective dielectric function. Subscripts *a* and *b* indicate the two in-plane directions of the monolayer. In each panel, the dielectric functions along the *a* and *b* directions are shown in red and blue, respectively.

# Supplementary Information For

# Hyperbolic Dispersion and Low-Frequency Plasmons in Electrides

Qi-Dong. Hao[1,2], Hao. Wang[1], Hong-Xing. Song[1], Xiang-Rong. Chen[2]*, Hua-Yun. Geng[1,3]*

[1] *National Key Laboratory of Shock Wave and Detonation Physics, Institute of Fluid Physics, China Academy of Engineering Physics, Mianyang, Sichuan 621900, P. R. China;*

[2] *College of Physics, Sichuan University, Chengdu 610065, P. R. China;*

3 *HEDPS, Center for Applied Physics and Technology, and College of Engineering, Peking University, Beijing 100871, P. R. China;*

\* *Corresponding authors. E-mail: s102genghy@caep.cn; xrchen@scu.edu.cn*

# Supplementary Figures

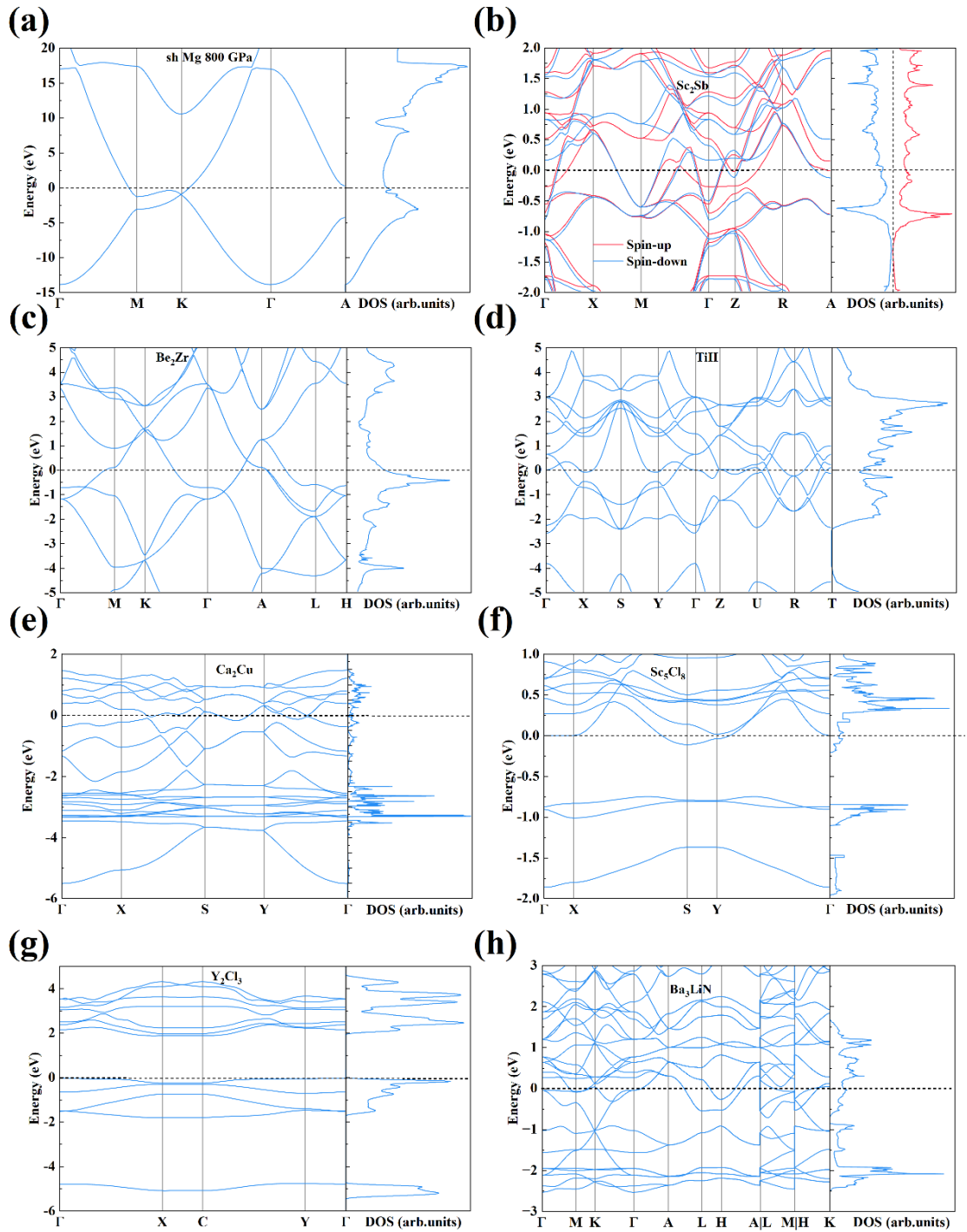

**Fig. S1.** Calculated band structures and density of states of (a) *sh* Mg at 800 GPa, (b)Sc$_2$Sb, (c)Be$_2$Zr, (d)TiH, (e)Ca$_2$Cu, (f)Sc$_5$Cl8, (g)Y$_2$Cl$_3$, (h)Ba$_3$LiN, respectively.

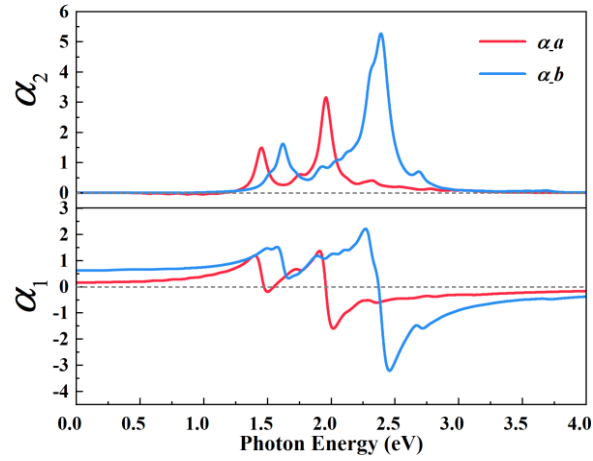

**Fig. S2.** Electronic polarizability of $Y_2Cl_3$.

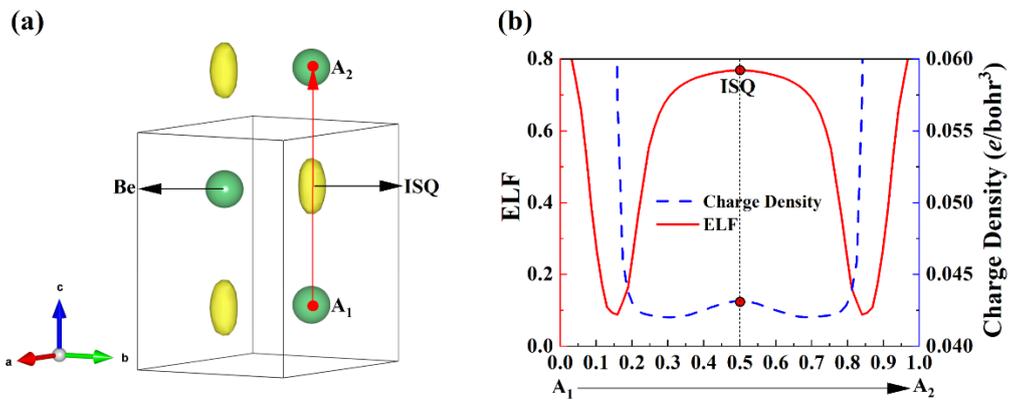

**Fig. S3.** (a) Crystal structure of *hcp* Be and its ELF at ambient pressure, plotted with an isosurface of 0.75. (b) Variation of ELF and charge density along the indicated path from atom $A_1$ to atom $A_2$ at ambient pressure; the path midpoint corresponds to the interstitial charge localization site.

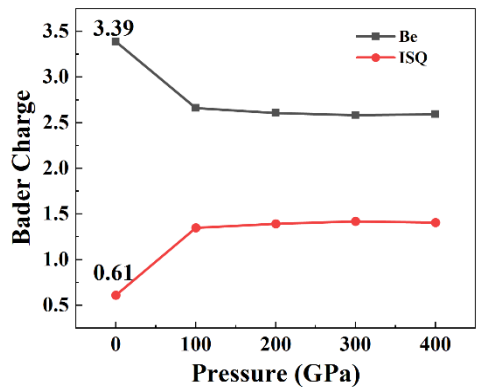

**Fig. S4.** Bader charges of each Be atom and ISQ in *hcp* Be under different pressures.

# Supplementary Note 1

As shown in Fig. S3(a), the *hcp* Be crystal features two Be atoms on Wyckoff 2c sites, while the symmetrically related Wyckoff 2d interstitial sites—located at the centers of polyhedral voids formed by five surrounding Be atoms—exhibit pronounced electron localization. The electron localization function (ELF) at these 2d sites peaks above 0.75, significantly higher than the free-electron–gas benchmark of 0.5. This indicates a highly localized accumulation of electrons in the 2d voids, in stark contrast to the delocalized electron distribution typical of ordinary metals.

To elucidate the spatial distribution of these electrons, we examined the variation of ELF and charge density along the $A_1$–$A_2$ path in Fig. S3(a). Both quantities display a bimodal profile: maxima occur at the path endpoints (the $A_1$/$A_2$ nuclear positions), consistent with the classical picture of peak electron density at atomic cores, and a second pronounced peak appears at the midpoint (the 2d interstitial site). Such dual maxima are atypical: in conventional covalent systems, although electron pairing can raise the ELF at the bond center, the charge density there is usually a local minimum. Here, however, the 2d void simultaneously exhibits ELF > 0.75 and a local charge-density maximum, ruling out covalent bonding as the cause of localization and providing definitive evidence for an interstitial quasi-atom (ISQ).

It is worth noting that previous studies have shown elemental electrides generally require high pressures to stabilize their lowest-energy structures[22], a phenomenon extensively validated in alkali (Li, Na) and alkaline-earth (Mg) systems [59,64,73]. By contrast, *hcp* Be displays electride characteristics at ambient pressure. At 0 GPa, Bader analysis (Fig. S4) reveals that each ISQ carries 0.61 e⁻, indicating that approximately 61 % of Be's valence electrons ($2s^2$ in the ground state) have transferred into the interstitial sites. Under 100 GPa, the ISQ charge rises markedly to 1.33 e⁻, leaving each Be atom with only 2.67 e⁻. Assuming the $1s^2$ core remains inert under pressure, Be's effective valence electron count drops to just 0.67 e⁻—half that of the ISQ. Beyond 100 GPa, the ISQ charge shows pressure-independent behavior, suggesting that the charge-transfer process has approached saturation.

**Supplementary Table**

**Table S1.** Structural information of the optimized computational models

| Materials | Space group | Lattice parameters (Å) | Atom | Wyckoff site | Atomic coordinates (fractional) |
|---|---|---|---|---|---|
| **hcp Mg** | $P6_3/mmc$ | a = b = 3.190<br>c = 5.179<br>α = β = 90.000<br>γ = 120.000 | Mg | 2d | 0.667 0.333 0.250 |
| **sh Mg** | $P6/mmm$ | a = b = 3.050<br>c = 3.045<br>α = β = 90.000<br>γ = 120.000 | Mg | 1a | 0.000 0.000 0.000 |
| **sh Mg (800 GPa)** | $P6/mmm$ | a = b = 1.882<br>c = 1.697<br>α = β = 90.000<br>γ = 120.000 | Mg | 1a | 0.000 0.000 0.000 |
| **hcp Be** | $P6_3/mmc$ | a = b = 2.265<br>c = 3.567<br>α = β = 90.000<br>γ = 120.000 | Be | 2c | 0.667 0.333 0.750 |
| **hp4 Na (230 GPa)-** | $P6_3/mmc$ | a = b = 2.860<br>c = 4.300<br>α = β = 90.000<br>γ = 120.000 | Na | 2a<br>2c | 0.000 0.000 0.000<br>0.333 0.667 0.250 |
| **TiH** | $P4_2/mmc$ | a = b = 2.948<br>c = 4.568<br>α = β = γ = 90.000 | Ti | 2c | 0.500 0.000 0.500 |
|  |  |  | H | 2e | 0.000 0.000 0.250 |
| **Sc₂Sb** | $P4/nmm$ | a = b = 4.208<br>c = 7.825<br>α = β = γ = 90.000 | Sc | 2a<br>2c | 0.000 0.000 0.000<br>0.658 0.829 0.250 |
|  |  |  | Sb | 2c | 0.500 1.000 0.720 |
| **Be₂Zr** | $P6/mmm$ | a = b = 3.812<br>c = 3.246<br>α = β = 90.000<br>γ = 120.000 | Be | 2d | 0.667 0.333 0.500 |
|  |  |  | Zr | 1a | 0.000 0.000 0.000 |
| **Ba₃LiN** | $P6_3/mmc$ | a = b = 8.289<br>c = 6.942<br>α = β = 90.000<br>γ = 120.000 | Ba | 6h | 0.148 0.852 0.750 |
|  |  |  | Li | 2c | 0.333 0.667 0.250 |
|  |  |  | N | 2a | 0.000 0.000 0.000 |
| **Monolayer-Sc₅Cl₈** | $P2/m$ | a = 3.534<br>b = 13.424<br>c = 25.192<br>α = β = γ = 90.000 | Sc | 1d<br>2n<br>2m | 0.000 0.000 0.500<br>0.500 0.334 0.507<br>1.000 0.509 0.559 |
|  |  |  | Cl | 2n | 0.5000 0.392 0.603 |
| | | | | 2n | 0.5000 0.017 0.431 |
| | | | | 2n | 0.5000 0.639 0.592 |
| | | | | 2m | 1.000 0.806 0.486 |
| **Monolayer-Ca$_2$Cu** | *P2$_1$/m* | a = 5.942 | Ca | 2e | 0.869 0.250 0.551 |
| | | b = 4.204 | | 2e | 0.373 0.750 0.596 |
| | | c = 25.192 | Cu | 2e | 0.386 0.250 0.512 |
| | | α = β = γ = 22.000 | | | |
| **Monolayer-Y$_2$Cl$_3$** | *C2/m* | a = 18.652 | Y | 4i | 0.643 0.504 1.000 |
| | | b = 25.257 | | 4i | 0.496 0.564 0.500 |
| | | c = 3.7934 | Cl | 4i | 0.587 0.605 1.000 |
| | | α = β = γ = 90.000 | | 4i | 0.401 0.600 1.000 |
| | | | | 4i | 0.497 0.564 0.500 |

## Supplementary References

1   Miao, M. S. & Hoffmann, R. High pressure electrides: a predictive chemical and physical theory. *Acc. Chem. Res.* **47**, 1311-1317 (2014).
2   Ma, Y. *et al.* Transparent dense sodium. *Nature* **458**, 182-185 (2009).
3   Pickard, C. J. & Needs, R. Dense low-coordination phases of lithium. *Phys. Rev. Lett.* **102**, 146401 (2009).
4   Miao, M. S. & Hoffmann, R. High-pressure electrides: the chemical nature of interstitial quasiatoms. *J. Am. Chem. Soc.* **137**, 3631-3637 (2015).